# Using Modularized Pin Ridge Filter in Proton FLASH Planning for Liver Stereotactic Ablative Body Radiotherapy

Chaoqiong Ma, Xiaofeng Yang[*], Yinan Wang, David Yu, Pretesh Patel, and Jun Zhou[*]

Department of Radiation Oncology and Winship Cancer Institute,

Emory University, Atlanta, GA, 30322, USA

*Email: xiaofeng.yang@emory.edu and jun.zhou@emory.edu

**Abstract**

We previously developed a FLASH planning framework for streamlined pin-ridge-filter (pin-RF) design, demonstrating its feasibility for single-energy proton FLASH planning. In this study, we refined the pin-RF design for easy assembly using reusable modules, focusing on its application in liver stereotactic ablative body radiotherapy (SABR). This framework generates an intermediate intensity-modulated proton therapy (IMPT) plan and translates it into step widths and thicknesses of pin-RFs for a single-energy FLASH plan. Parameters like energy spacing, monitor unit limit, and spot quantity were adjusted during IMPT planning, resulting in pin-RFs assembled using predefined modules with widths from 1 to 6 mm, each with a WET of 5 mm. This approach was validated on three liver SABR cases. FLASH doses, quantified using the FLASH effectiveness model at 1 to 5 Gy thresholds, were compared to conventional IMPT (IMPT-CONV) doses to assess clinical benefits. The highest demand for 6 mm width modules, moderate for 2-4 mm, and minimal for 1- and 5-mm modules were shown across all cases. At lower dose thresholds, the two-beam case showed significant dose reductions (>23%), while the other two three-beam cases showed moderate reductions (up to 14.7%), indicating the need for higher fractional beam doses for an enhanced FLASH effect. Positive clinical benefits were seen only in the two-beam case at the 5 Gy threshold. At the 1 Gy threshold, the FLASH plan of the two-beam case outperformed its IMPT-CONV plan, reducing dose indicators by up to 28.3%. However, the three-beam cases showed negative clinical benefits at the 1 Gy threshold, with some dose indicators increasing by up to 16% due to lower fractional beam doses and closer beam arrangements. This study evaluated the feasibility of modularizing streamlined pin-RFs in single-energy proton FLASH planning for liver SABR, offering guidance on optimal module composition and strategies to enhance FLASH planning.

## 1. Introduction

FLASH radiotherapy (FLASH-RT), employing ultrahigh-dose-rate irradiation typically exceeding 40 Gy/s, has emerged as a promising approach for reducing toxicity in healthy tissues while effectively maintaining tumor control, surpassing conventional-dose-rate irradiation.(Favaudon *et al* 2014, Vozenin *et al* 2019, Fouillade *et al* 2020, Bourhis *et al* 2019a) Through the use of proton, electron, or photon beams at ultrahigh dose rates, small animal studies have demonstrated significant protective effects on normal structures such as the lungs, brain, skin, and abdominal tissues,(Favaudon *et al* 2014, Vozenin *et al* 2019, Loo *et al* 2017, Montay-Gruel *et al* 2019, 2017) suggesting a potential for broad application in sparing healthy tissues during cancer treatment. The underlying mechanisms of FLASH-RT, potentially linked to oxygen depletion and the production of reactive oxygen species, continue to be a focal point of ongoing research investigations.(Wilson *et al* 2020, Spitz *et al* 2019) The clinical utility of FLASH-RT was initially confirmed in a patient with T-cell cutaneous lymphoma, where electron beam treatment improved outcomes for both the tumor and surrounding skin.(Bourhis *et al* 2019b) Further clinical validation was provided by the FAST-01 trial, which demonstrated the efficacy of proton-based FLASH-RT in patients with multiple bone metastases.(Mascia *et al* 2023, Daugherty *et al* 2023) Ongoing trials continue to explore the use of proton therapy for chest bone metastases(Daugherty *et al* 2024a) and electron therapy for metastatic skin melanoma,(Anon n.d.) underpinning the expanding therapeutic scope of FLASH-RT.

Benefiting from the narrow Bragg peaks (BPs), proton beams offer superior precision in tumor targeting and normal tissue sparing when compared with photon and electron beams. This precision is enhanced by the widely used delivery method of pencil beam scanning (PBS) for intensity-modulated proton therapy (IMPT), which ensures high dose conformity to the target.(Newhauser and Zhang 2015) These attributes, coupled with the extensive treatment range of proton beams, establish them as the preferred modality for FLASH-RT. However, delivering proton beams at FLASH dose rates using a clinical cyclotron presents challenges due to the modulation of proton energies by an energy degrader and energy selection system, required to achieve spread-out Bragg peaks (SOBPs) for PBS-IMPT, significantly reducing beam current and impeding FLASH rate delivery.(Jolly *et al* 2020) To address this issue, preliminary studies have explored using the highest energy proton transmission beams (TBs) from a cyclotron for FLASH planning, utilizing the entrance dose region of the Bragg curves for target dose coverage while placing the BPs outside the patient.(Gao *et al* 2020, van de Water *et al* 2019, van Marlen *et al* 2020, Folkerts *et al* 2020) This method, however, risks increased exposure of normal tissue at the distal edge of the target to high exit dose of TBs. To overcome this, strategies have been developed to use the BPs of the highest energy for direct target coverage in FLASH planning, in which range compensators are customized to align the single-energy BPs to the distal edge of the target.(Wei *et al* 2022b, 2022a, Kang *et al* 2022) Additionally, efforts have been made to integrate BPs of conventional dose rates,(Lin *et al* 2021) or single-energy SOBPs created by shooting the highest energy proton beams through general bar ridge filters (RFs),(Ma *et al* 2023) with TBs for FLASH planning.

To maximize the utility of BPs in FLASH-RT, studies has focused on the use of pin-shaped RFs (pin-RFs), which consist of step-shaped ridge pins designed to produce a single-energy SOBP along each pencil beam direction (PBD) for IMPT-PBS.(Simeonov *et al* 2017, Schwarz *et al* 2022, Liu *et al* 2023, Zhang *et al* 2022a) These pin-RFs are tailored primarily in two ways for single-energy proton FLASH planning. The first approach involves designing the step widths and thicknesses of each ridge pin to create a uniform SOBP that conforms to the target along the corresponding PBD, proving feasible for FLASH therapy by simultaneously accounting for dose, dose rate coverage, and linear energy transfer (LET) effects.(Simeonov *et al* 2017, Schwarz *et al* 2022, Liu *et al* 2023) Alternatively, ridge pins can be configured to modulate the dose distribution along each PBD in a manner similar to traditional PBS-IMPT, allowing for more flexible manipulation of dose distributions.(Zhang *et al* 2022a) However, these pin-RF designs typically feature more than 10 steps per pin with a fine resolution of 0.1 mm, necessitating a high-precision and time-intensive production process, such as 3D printing. Building on the second design technique, a streamlined pin-RF design method was developed in our previous study, creating pin-RFs with coarser resolution suitable for FLASH planning.(Ma *et al* 2024) This innovation potentially allows for more efficient production, the possibility of assembling with reusable modules, and thus offers flexible adjustability and reduced production costs for the pin-RFs.

Proton stereotactic ablative body radiotherapy (SABR) has demonstrated superior prognostic outcomes in the treatment of liver cancer.(Nabavizadeh *et al* 2018) With the integration of FLASH delivery, proton SABR for liver cancers offers the potential to reduce normal tissue toxicities thereby enhancing therapeutic efficacy.(Levy *et al* 2020) Leveraging our previously developed FLASH planning framework for streamlined pin-RF design, this study aimed to further refine the streamlined pin-RF design and explore its modularization using reusable modules. We specifically focused on applying this approach to FLASH planning for liver SABR, with the goal of assessing the achievable FLASH effect using this refined planning strategy.

## 2. Methods and materials

The following subsections begins with an overview of our in-house FLASH planning framework for the streamlined pin-RF design, and its implementation in this study. Following this, we elaborate on the modularization strategy for the streamlined pin-RFs, highlighting the selection of specific parameters in the FLASH planning framework and the design of reusable unit modules for efficient pin-RF assembly. In the context of implementing these modularized pin-RFs for liver SABR FLASH planning, we further provide details on adopting FLASH effectiveness model to assess the FLASH effect of the FLASH plans for the liver cases.

## 2.1 Streamlined pin-RF design for FLASH planning

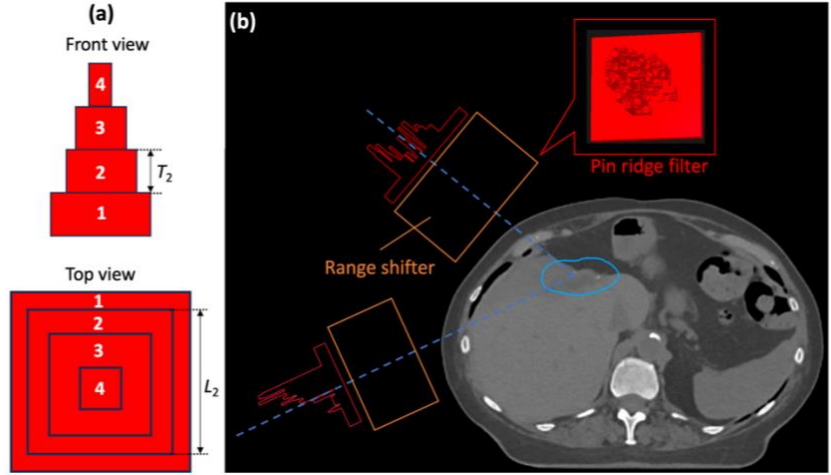

**Figure 1.** (a) Front and top views of a four-step ridge pin. $T_2$ and $L_2$ are the thickness and width of the second step from the bottom, respectively. (b) A two-field arrangement for a single-energy proton FLASH plan using designed pin ridge filters, with a 3D view of one pin RF. The target is outlined in blue.

Previously, we proposed a streamlined pin-RF designing approach for single-energy proton FLASH planning.(Ma *et al* 2024) This method is based on the concept of converting pin-RFs from an IMPT plan,(Zhang *et al* 2022b) where each pyramid-shaped ridge pin featuring variable step thicknesses and surface areas is used to modulate the depth dose curve of a monoenergetic proton beam along a specific PBD to replicate the depth dose curve along the same PBD in the IMPT plan. Figure 1(a) displays the front and top views of a four-step ridge pin. The water equivalent thickness (WET), $T_i$, of step $i$ ($i >1$) in a ridge pin, can be calculated from the BP depths, $R_i$ and $R_{i-1}$, of the $i^{th}$ and $(i-1)^{th}$ highest energies, respectively, along the corresponding PBD in the IMPT plan, using equation (1). And WET of the bottom step can be derived the using equation (2), which factors in the WETs of the pin-RF base and the range shifter (RS), $T_{RS}$ and $T_B$, respectively, for the given field. Figure 1(b) represents a two-field setup of the pin-RFs and RSs for a liver case, where the RS associated with each field serves to pull up the BPs of the highest energy beam with a range of $R_0$ (e.g., 250 MeV of a Varian ProBeam at our facility) to the distal edge of the target, thus allowing the corresponding pin-RF to spread out the BPs within the target. The step width, $L_i$, of step $i$ can be derived using equations (3) and (4), accounting for the relative weight, $\overline{w}_i$, of the $i^{th}$ highest energy among all the $N$ energies along the PBD in the IMPT plan, where $w_i$ denotes the weight of the $i^{th}$ highest energy along this PBD.

$$T_i = R_i - R_{i-1} \tag{1}$$

$$T_1 = R_0 - T_{RS} - T_B \tag{2}$$

$$\bar{w}_i = w_i / \sum_{i=1}^{N} w_i \tag{3}$$

$$\bar{w}_i = (L_{i-1}^2 - L_i^2)/L_1^2 \tag{4}$$

To design the streamlined pin-RFs featuring a coarse resolution and sparse pin distribution, we established an inverse planning framework which is integrated within a Treatment Planning System (TPS), RayStation 10B (RaySearch Laboratories, Stockholm, Sweden), for FLASH planning.(Ma *et al* 2024) This framework initiates by creating an IMPT plan which incorporates multiple energy layers generated through a downstream energy modulation strategy, referred to as IMPT-DS plan. It then proceeds with a nested pencil-beam-direction-based (PBD-based) spot reduction process to iteratively eliminate the low-weighted energy layers along each PBD and low-weighted PBDs from the IMPT-DS plan. Subsequently, the IMPT-DS plan is translated into the pin-RFs and the single-energy beam configurations, including spot positions and weights, for the FLASH plan (referred to as IMPT-RF). The pin-RFs produced through this approach are characterized by fewer ridge pins and reduced steps along each pin, facilitating efficient production.

In this study, we implemented this FLASH planning framework for liver SABR cases. Shifting from the use of PMMA in our previous study, we adopted graphite as the material for RSs, a common material used for energy degraders due to its reasonably low atomic number (Z = 6), good mechanical robustness, and cost-effectiveness.(Psoroulas *et al* 2020) Compared to PMMA, graphite exhibits a lower multiple Coulomb scattering (MCS) effect of protons. Furthermore, the higher relative stopping power (RSP) to water of graphite, compared to PMMA (1.68 vs. 1.17), allows for an over 40% reduction in the WET of the RS. This reduction can shorten the distance from the pin-RF to the isocenter (RF-to-iso), consequently diminishing the inverse square effect and enhancing the dose conformity of the IMPT-RF. By switching to graphite as the RS material, we were able to reduce the RF-to-iso distance from 31 cm to 26 cm in this study.

**2.2 Modularization of the pin-RFs**

The primary goal of this study is to produce the streamlined pin-RFs of pyramid-shaped ridge pins for liver SABR FLASH planning through a modular strategy. This involves assembling the pin-RFs with cuboid-shaped unit modules that have specified widths and WETs, chosen from a predetermined collection of reusable modules. As mentioned earlier, these pin-RFs are translated from an IMPT-DS plan in our FLASH planning process. Thus, the resolution and sparsity of the ridge pins, which determine the shapes and quantities of the required unit modules for assembly, respectively, can be adjusted through the parameter settings applied in this intermediate IMPT planning.

The resolution of the pyramid-shaped ridge pin is defined by the WET and width resolutions of the steps. The WET of each step indicates the variation in BP depths between two successive energies along the PBD in the corresponding IMPT-DS plan. Thus, the minimum WET across the steps of the ridge pin, reflecting the WET resolution of the ridge pin, matches the energy layer spacing adopted in the planning process for the IMPT-DS plan. This necessitates that the WET of the cuboid unit modules to assemble the ridge pins aligns with this minimum step WET. In this study, the energy layer spacing in the planning for the IMPT-DS plan was set to 5 mm, resulting in the WETs of the unit modules featuring a 5 mm resolution. In addition, the step width resolution of 1 mm of the ridge pin was aligned with that of the streamlined pin-RF from our previous implementation of the FLASH planning framework. Attaining this coarse resolution necessitates a strategic reduction in the number of steps per ridge pin, facilitated by the nested PBD-based spot reduction process in IMPT planning. This process is essential due to the step widths of a ridge pin are calculated from the spot weights along the corresponding PBD of the IMPT-DS plan, as outlined in equations (3) and (4). This is followed by rounding the calculated step widths to the nearest whole millimeter. A thorough explanation of achieving this resolution is provided in our previous study.(Ma *et al* 2024) Given that the bottom step width, $L_1$, of a ridge pin is 6 mm, we designed the unit modules in six specific widths, ranging from 1 to 6 mm in 1 mm increments. Consequently, this module set comprises of unit modules in six distinct sizes, each featuring a WET of 5 mm and widths ranging from 1 to 6 mm, in 1 mm intervals, as depicted in Figure 2. This design allows for the assembly of any step of a pin-RF by stacking unit modules with the same width as the step.

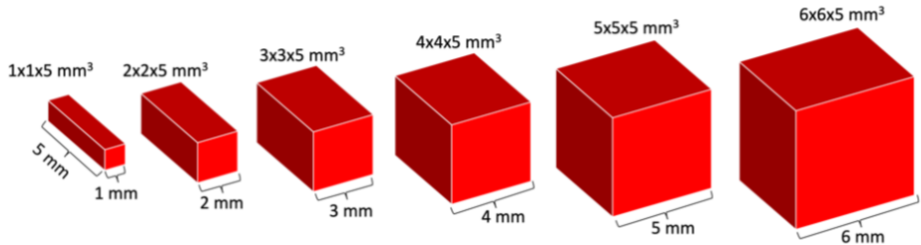

**Figure 2.** Unit modules of six sizes

Furthermore, reducing the total number of ridge pins is essential for efficiently assembling the pin-RFs. This can be accomplished by eliminating the low-weighted PBDs with total spot monitor units (MUs) below a predefined threshold during the nested PBD-based spot reduction process in the IMPT-DS planning. It is important to note that the spot MU for a PBD of the translated IMPT-RF corresponds to the cumulative spot MUs for that PBD in the IMPT plan. Therefore, this MU threshold should be sufficiently high to allow the minimum spot weight of the IMPT-RF to be deliverable at a high beam current, necessary for achieving FLASH dose rates. In this study, the MU threshold was set to 300 MU, ensuring the deliverability of the IMPT-RF at a beam current of 500 nA and a minimum spot duration time of 0.5 ms.

## 2.3 Patient study

This modularized pin-RF design for FLASH planning was validated on three liver cancer patients previously underwent conventional IMPT. For each patient, an IMPT-RF plan employing the modularized pin-RFs was generated. The sizes of clinical target volume (CTV) for Patient/Case 1, 2 and 3 were 48.9, 65.4 and 125.7 cc, respectively. The prescription dose was set to 50 Gy delivered in five fractions in accordance with our clinical liver SABR guidelines. To ensure target dose conformity and avoidance of the normal tissues, the beam angles from the original conventional IMPT plans (IMPT-CONV) were maintained in the FLASH planning: two beams for patient 1, and three beams for patient 2 and 3. The objective for CTV coverage was that 95% of the CTV receives the full prescription dose, $D_{95\%} = 50$ Gy. Additionally, the maximum dose, $D_{max}$, of the CTV was limited to 140% of the prescription dose ($D_{max} \leq 70$ Gy). Robust optimization was incorporated into the IMPT-DS planning of the FLASH planning process, accommodating setup and range uncertainties of 5 mm and 3.5%, in alignment with our clinical protocol guidelines. Besides, single-field optimization (SFO) was utilized to achieve robust uniform field doses.

The quality of the IMPT-RF plan was assessed based on the concerned dose indicators for the normal tissues and the target dose conformity. The latter was quantified using the conformity index (CI), calculated as the ratio of the total volume that received the prescription dose to the CTV volume. Additionally, a robustness analysis was conducted for each IMPT-DS plan, which included 12 perturbation scenarios accounting for setup uncertainties in all three directions, along with a ±3.5% range uncertainty. Each scenario combined a setup error in one direction with a single-sign range uncertainty.

## 2.4 Evaluation of FLASH effect

The FLASH dose of the IMPT-RF plan was quantified using the FLASH effectiveness model.(Krieger *et al* 2022) This model operates on the principle that the FLASH effect is triggered in a normal tissue voxel if the dose ($\Delta d$) delivered within a time window ($\Delta t$) exceeds a dose threshold ($D_0$), $\Delta d \geq D_0$, and is delivered at an average dose rate surpasses a dose rate threshold ($\dot{D}_0$), $\Delta d/\Delta t \geq \dot{D}_0$. This FLASH effect remains active throughout the triggering window and an additional persistence period. The dose delivered with the FLASH effect is deemed biologically less "effective", and therefore, it is adjusted by an effective factor (<1), reflecting reduced toxicity.(Mazal *et al* 2020)

In this work, $\dot{D}_0$, persistence time and effective factor were set to 40 Gy/s, 200 ms and 0.67, respectively, mirroring the values from the aforementioned study.(Krieger *et al* 2022) The effective factor of 0.67 represents a 33% reduction in the dose delivered to normal tissues via active FLASH effect. Given the significant dependency of the FLASH effect on $D_0$, as demonstrated in the previous study,(Krieger *et al* 2022) $D_0$ of 1, 3 and 5 Gy were selected to evaluate the FLASH effective doses for each FLASH plan.

These $D_0$ settings, not exceeding the average beam dose per fraction (approximately 5 Gy for the two-beam plan) in the liver SABR planning, were chosen to ensure activation of the FLASH effect. Besides, the scanning pattern was considered, and a scanning speed of 10 mm/ms, as assumed by the Varian research group,(Folkerts *et al* 2020) was employed in evaluating the FLASH effective dose.

The clinical benefit of induced by FLASH was assessed by comparing the FLASH dose of IMPT-RF to the dose of an IMPT-CONV plan in the concerned normal tissues, and quantifying the achieved dose reduction. Since tumor cells are deemed un affected by FLASH, no dose reduction is applied within the gross tumor volume (GTV). In contrast, the margin extending from the GTV to planning target volume (PTV), known as PTV-GTV, is especially intriguing. This region, which potentially comprises both healthy tissue and tumor cells and is exposed to high doses of radiation, may exhibit a clinically relevant differential response if the FLASH effect operates at the cellular level.(Krieger *et al* 2022) Thus, the clinical benefits of FLASH in PTV-GTV was evaluated, along with its impact on other surrounding normal tissues. Furthermore, the impact of the number of beams on the FLASH effect was assessed by comparing the clinical benefits achieved by FLASH between Case 1, who was treated with two beams, and Cases 2 and 3, each treated with three beams in their FLASH planning.

## 3. Results

### 3.1 Analysis of the modularized pin-RFs

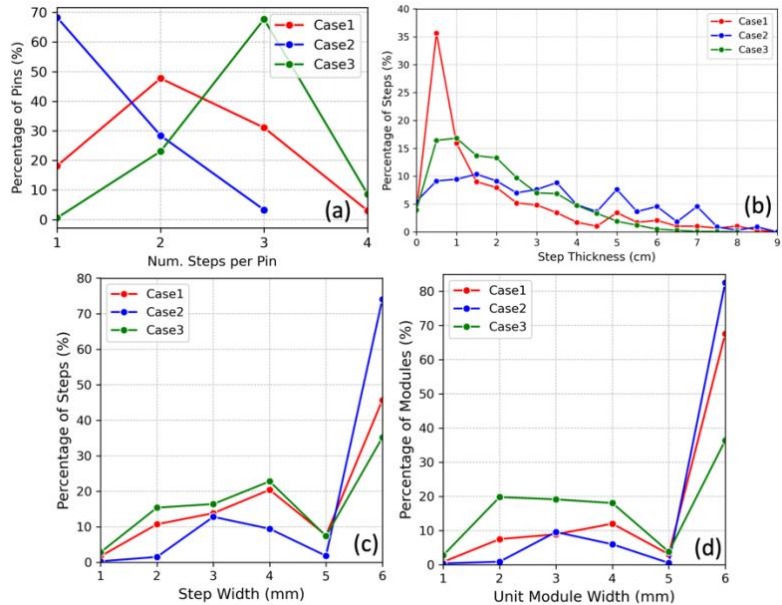

**Figure 3.** Frequency distributions for Cases 1 (red), 2 (blue), and 3 (green); (a) Percentage of pins by number of steps per pin, (b) Percentage of steps by WET, (c) Percentage of steps by width, and (d) Percentage of unit modules by width.

The total number of ridge pins in the pin-RFs of an IMPT-RF plan was proportional to the target size, with the highest count of 347 observed in Case 3 and the lowest of 132 in Case 1. A statistical analysis on the step counts per ridge pin revealed significant differences across the cases, with relative frequencies depicted in Figure 3(a) and absolute frequency distributions in Figure S1. Cases 1 and 3 featured ridge pins with up to four steps, whereas Case 2 was limited to a maximum of three steps per pin. Specifically, 68% of pins in Case 2 comprised only one step, in contrast to Case 3, where 68% contained three steps. This discrepancy likely raised from differing modulation complexities due to target size, with Case 3 requiring more complex energy modulation. Despite having the smallest target, Case 1 showed a higher energy modulation complexity than Case 2, with 79% of its pins having two or three steps, possibly due to fewer beams requiring more complex modulation per beam.

Figure 3(b) displays the relative frequency distribution of step WETs across pins for each case, with absolute distributions provided in Figure S1. The total number of steps in the pin-RFs for an IMPT-RF plan increased with target size; Case 1 had the fewest at 289, and Case 3 the most at 987. Step WETs typically reached up to 8-9 cm across all cases. Step WETs across all cases reached up to 8-9 cm, but were predominantly below 3.5 cm in Cases 1 and 3, with approximately 87% of steps within this range. Additionally, the most common WET was 0.5 cm in Case 3, representing 35% of steps. In contrast, Case 2 exhibited a more uniform distribution with 67% of steps thinner than 3.5 cm. The frequent occurrence of thinner steps in Cases 1 and 3, relative to Case 2, indicated a finer energy modulation in these cases, as thinner steps correlate with smaller energy discrepancies along each PBD. Note that the bottom step of some ridge pins, at 0 cm thickness, corresponded to the highest energy used in IMPT-DS planning. Furthermore, the relative and absolute frequency distributions of step widths across pins for each case were analyzed, as shown in Figures 3(c) and S1. All pins started with a base step width of 6 mm, which was the most common across all cases, matching the total number of pins per case. The proportion of this step width was inversely related to the total step count, peaking at 74% in Case 2. Conversely, step widths of 1, 2, and 5 mm were less frequent in Case 2, each comprising less than 2%, while in Cases 1 and 3, widths of 1 and 2 mm accounted for less than 8% each.

We conducted an analysis of the demand for unit modules of varying widths required for pin-RF assembly, which involves stacking unit modules corresponding to the widths of steps in each pin. As depicted in Figure 3(d), the demand pattern for unit module widths mirrors the frequency distribution of step widths for each case shown in Figure 3(c). Modules of widths 6 mm were in highest demand, accounting for 67%, 83%, and 36% of total modules in Cases 1, 2, and 3, respectively. Despite Case 3 requiring the highest number of 6 mm wide steps, Case 2 demanded the greatest quantity of unit modules of widths 6 mm (1634 compared to 1425 for Case 3), due to generally thicker bottom steps as shown in Figure S1. Case 1, with the smallest target size, required the fewest 6 mm modules at 719. There was also notable demand for modules of widths 2 to 4 mm in Cases 1 and 3, and 3 and 4 mm in Case 2, with average demands of 101 (9%), 154 (7%), and 745 (19%) for Cases 1, 2 and 3, respectively. The least demanded widths across all cases were 1 and 5 mm, as well as 2 mm for Case 2, each with a relative frequency below 4%. Therefore, this analysis could inform the optimal composition of the module set, primarily featuring unit modules of 6 mm width, supplemented by a balanced selection of modules ranging from 2 to 4 mm in width, and a minimal quantity of modules of widths 1 and 5 mm.

### 3.2 Dosimetric assessment of FLASH planning

For each case, the translated IMPT-RF plan from the IMPT-DS plan met all dosimetric criteria, with the normalization ensuring that 95% of the CTV received the prescribed dose. The plan quality of the IMPT-DS plan was well preserved by the corresponding IMPT-RF plan. Figure 4 (a) and (b) depict the 2D dose distributions of the IMPT-DS and IMPT-RF plans for Case 1, respectively, serving as an example. Notably, there was a minor increase in the low-dose spillage near the distal edge of the target in the IMPT-RF plan, attributed to the increased spot size/penumbra from the pin-RFs. Additionally, relevant dose indicators of the concerned surrounding normal tissues were derived from the dose volume histograms (DVHs) of both plans across all cases, which are presented in Table S1-S3 in the Supplementary Material. Illustratively, Figure 4 (c) presents the DVHs comparison between the two plans for Case 1.

Specially, for each case, a negligible CI discrepancy of 0.02 or less was noted between the two plans, indicating comparable target dose conformity. Regarding normal tissue sparing, variations in the dose indicators representing maximum doses for a point or small volume, such as the maximum dose ($D_{max}$) of skin across all cases and $D_{5cc}$ of rib for Case 2, generally remained marginal, not exceeding 3.5 Gy between the two plans. The exception was the $D_{max}$ of esophagus for Case 2, which experienced a notable increase of up to 6.7 Gy from the IMPT-DS to the IMPT-RF plan. This significant rise was due to the esophagus being located in a relatively low-dose region, making it more susceptible to the increased low-dose spillage caused by the enlarged spots from the pin-RFs. It is noteworthy that increase in $V_{21Gy}$ of liver excluding the GTV (liver-GTV), measuring the volume of liver-GTV receiving at least 21 Gy, from IMPT-DS to IMPT-RF was minimal (approximately 1%) for Case 1. In contrast, this metric increased by 8.6-10.2% for Cases 2 and 3. This relatively large rise in $V_{21Gy}$ for Cases 2 and 3, compared to Case 1, can be attributed to the use of an additional beam in the planning. More beams typically lead to greater low-dose exposure, particularly when additional pin-RFs are employed, which contribute to increased low-dose spillage. Consequently, this exposes a larger volume of the liver to lower doses, leading to an increase in $V_{21Gy}$ of liver-GTV for Cases 2 and 3.

Results from the robustness analysis are shown in Figure 4 (d), which features DVHs for both the nominal dose and 12 perturbed doses of the IMPT-DS plan for Case 1, provided as an illustrative example. We assessed plan robustness using the $D_{95\%}$ of CTV from the second worst-case scenario, with values ranging from 84.0-88.5% across the three cases; the lowest was observed in Case 1. The reduced robustness of the IMPT-DS plan for Case 1 can be attributed to two factors: the use of fewer beams in planning, which compromises robustness of the plan against setup uncertainties, and the deeper location of the target relative to the other cases, which introduces greater range uncertainties. Notably, significant discrepancies between the perturbed and nominal DVHs for the duodenum are evident in Figure 4 (d), attributable to its small volume and proximity to the target. Although the $D_{95\%}$ of CTV for each case in the second worst-case scenario fell below the prescribed dose, it is important to note that scaling up the nominal dose could safely elevate it to the 95% threshold, in line with plan objectives. Hence, the IMPT-DS plan is capable of maintaining adequate target coverage despite the setup and range uncertainties, thereby ensuring the robustness of the translated IMPT-RF plan for FLASH delivery.

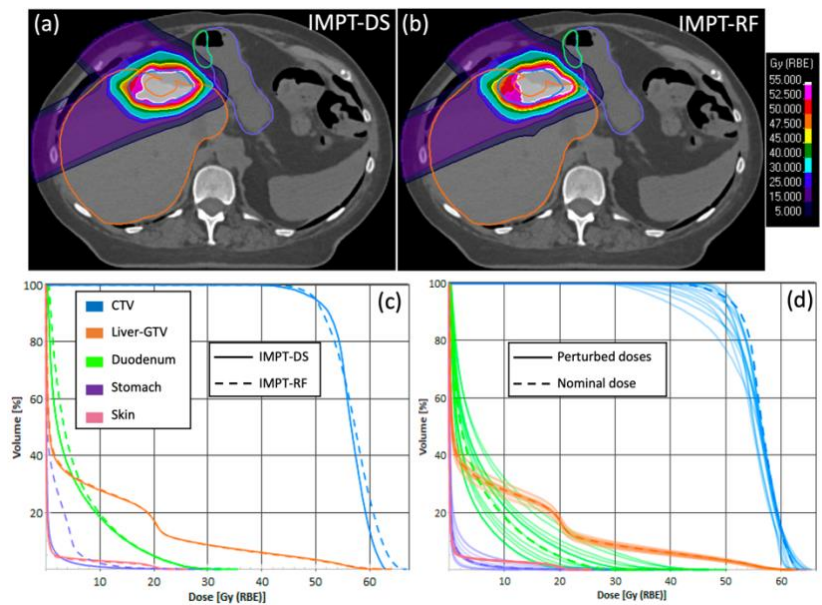

**Figure 4.** Dosimetric Analysis for Case 1. 2D dose distributions for (a) IMPT-DS and (b) IMPT-RF plans. (c) DVH comparisons between IMPT-DS and IMPT-RF. (d) DVHs of 12 perturbed doses and the nominal dose. The CTV, liver-GTV, duodenum, stomach and skin are contoured in blue, orange, green, purple and pink, respectively. The dose scale in isodose plot ranges from 10% to 110% of the prescription dose.

### 3.3 Evaluation of the FLASH effect

The FLASH effect of the IMPT-RF plan with modularized pin-RFs were assessed by comparing the FLASH doses of the IMPT-RF plan at various dose threshold $D_0$ of 5, 3 and 1 Gy, referred to as FLASH-5Gy, FLASH-3Gy, and FLASH-1Gy, respectively, with the dose of the IMPT-RF plan. Figure 5 illustrates the results for Case 1, and the results for the other two are available in Figures S2 and S3 in the Supplementary Material. Additionally, dose indicators for relevant normal structures, including PTV-GTV for all cases, are detailed in Tables S1 to S3.

Identical dose distributions were noted between the FLASH-5Gy and the IMPT-RF plans for Cases 2 and 3, as the $D_0$ of 5 Gy exceeded the fractional beam doses in these three-beam plans. With the integration of SFO into the FLASH planning process, each beam in a three-beam plan delivered about 3.3 Gy to the target in each fraction. Consequently, at a $D_0$ of 5 Gy, a FLASH effect could not be initiated for the three-beam plans. In contrast, the FLASH effect was activated for Case 1 of the two-beam plans, leading to a significant 23.2% reduction in $D_{mean}$ of PTV-GTV and a slight reduction in $D_{max}$ of skin. However, despite the $D_{max}$ of stomach exceeding 25 Gy (> 5 Gy per fraction), no reduction was observed with FLASH at a $D_0$ of 5 Gy. This was attributed to the anatomical positioning, where the maximum dose occurred at the intersection of two beams, and individual beam doses were insufficient to trigger the FLASH effect, unlike the skin where the maximum dose was delivered along a single beam path.

Lowering $D_0$ to 3 Gy resulted in significant dose reductions for Case 1, with FLASH-3Gy showing decreases from 25.6 to 31.5% in $D_{max}$ of skin, $V_{21Gy}$ of liver-GTV, and $D_{mean}$ of PTV-GTV compared to IMPT-RF. However, for Cases 2 and 3, reductions were limited to 14.7% in these indicators. This diminished FLASH effect in Cases 2 and 3 can be attributed to lower fractional beam doses, which shortened the FLASH triggering windows and resulted in lesser dose reductions. Notably, reductions in these three dose indicators were lower for Case 3 compared to Case 2, with no reductions observed in $D_{max}$ of skin and $V_{21Gy}$ of liver-GTV for Case 3. This difference can be attributed to the closer beam arrangements in Case 3, which led to a larger volume of normal tissues at the beam intersections receiving high doses (> 45 Gy). Consequently, the fractional beam doses were either insufficient to trigger FLASH or resulted in less FLASH effect for Case 3. Similar trend was noted in other dose indicators across the cases, highlighting that delivering a high fractional beam dose to a specific area is essential for activating the FLASH effect in the normal tissues within that region.

Reducing $D_0$ from 3 to 1 Gy led to significant reductions in dose indicators across the cases: 8.3 to 23.2% in $V_{21Gy}$ of liver-GTV and $D_{max}$ of stomach in Case 1, $D_{max}$ of great vessels and esophagus in Case 2, and $D_{max}$ of duodenum, skin, and $D_{0.5cc}$ of bowel bag in Case 3, comparing FLASH-3Gy to FLASH-1Gy. However, minimal or no reductions—less than 5%—were seen in other dose indicators when lowering $D_0$ from 3 to 1 Gy, such as the unchanged $D_{max}$ of skin for Case 2, as illustrated in Figure S4 of the Supplementary Material. Although some dose reductions in skin from FLASH-3Gy to FLASH-1Gy for Case 2 were observed, certain areas receiving significant beam doses showed no reduction. This could be due to the accumulation of doses from multiple spot deliveries, resulting in an insufficient dose rate to trigger the FLASH effect.

### 3.4 Clinical benefits brought by FLASH

Comparisons between the FLASH doses and the reference doses from the IMPT-CONV plan were conducted to evaluate the clinical advantages of FLASH across various cases. Results for Case 1 are visually depicted in Figure 5, while results for Cases 2 and 3 are available in Figures S2 and S3 in the Supplementary Material. Dose indicators for normal structures related to the IMPT-CONV plans are detailed in Tables S1-S3. Since no FLASH effect was obtained at $D_0$ of 5 Gy for Cases 2 and 3, these comparisons were excluded. Comparing FLASH-5Gy to IMPT-CONV for Case 1, a significant reduction of 19.4% in $D_{mean}$ of PTV-GTV was observed, suggesting a positive clinical benefit on potential normal tissues in this region. Conversely, other dose indicators experienced increases of up to 43.7%, with $V_{21Gy}$ of liver-GTV showing the highest increase, indicating negative clinical benefits induced to these normal tissues by the FLASH plan.

Lowering $D_0$ to 3 Gy for Case 1 resulted in enhanced clinical benefits, notably achieving a substantial 28.1% reduction in $D_{mean}$ of PTV-GTV in FLASH-3Gy compared to IMPT-CONV. Positive clinical benefits were observed across most normal tissues, with only a moderate 9.6% increase in $D_{max}$ of the stomach noted. Particularly, this adjustment nearly eliminated the substantial negative clinical impact previously seen in liver-GTV at $D_0$ of 5 Gy. In contrast, Case 2 and 3 showed reductions of only 16.0% and 4.1% in $D_{mean}$ of PTV-GTV, respectively, along with a moderate 6.3% reduction in $D_{max}$ of rib for Case 2. Negative clinical benefits were observed in other normal tissues when comparing FLASH-3Gy to IMPT-CONV for these two cases. As mentioned earlier, this discrepancy in clinical benefit compared to Case 1 could result from the reduced fractional beam doses due to additional beam or closer beam arrangement. This effect may be amplified in regions experiencing low-dose spillage due to enlarged spots in FLASH plan delivery, potentially leading to significant negative clinical benefits. For example, an increase of up to 10.1 Gy in $D_{max}$ of the esophagus was observed for Case 2.

At $D_0$ of 1 Gy, Case 1 showed positive clinical benefits across all concerned normal tissues, with reductions in relevant dose indicators ranging from 4.8% to 28.3% when comparing FLASH-1Gy to IMPT-CONV. Conversely, Cases 2 and 3 displayed negative clinical benefits on certain normal tissues. For example, there were increases of about 16% in the maximum skin dose for Case 2 and in the liver-GTV $V_{21Gy}$ for Case 3. It is noteworthy that reducing $D_0$ from 3 to 1 Gy still led to some significant negative clinical benefits, such as an 8.8 Gy increase in Dmax of esophagus for Case 2. However, drastically enhanced clinical benefits were obtained for Case 3, including a reduction of 14.5% in Dmax of duodenum and an almost unchanged $D_{0.5cc}$ of the bowel bag when comparing FLASH-1Gy to IMPT-CONV. Normal tissues subjected to high fractional beam doses, such as the bowel bag in Case 3, or situated in high-dose regions minimally impacted by the low-dose spillage from enlarged spots in FLASH plan delivery—for example, PTV-GTV of all cases and the duodenum of Case 3—tend to have extended FLASH triggering windows with reduced $D_0$. These tissues generally gain more from the FLASH effect with reduced $D_0$, leading to either positive or minimally negative clinical outcomes. These tissues typically gain more from the FLASH effect, resulting in positive or minimal negative clinical benefits.

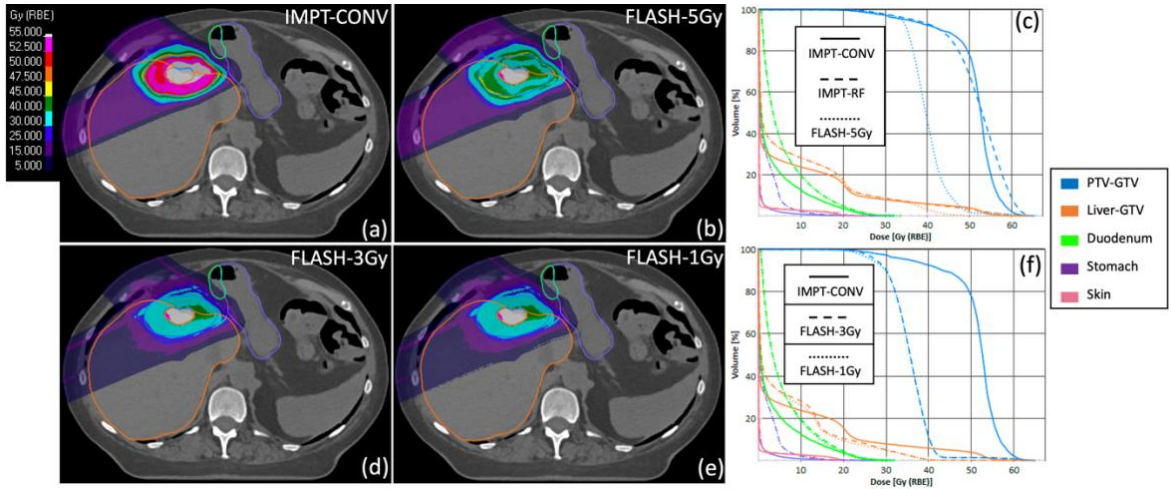

**Figure 5.** FLASH assessment for Case 1. 2D dose distributions for (a) IMPT-CONV and (b) FLASH-5Gy. (c) DVH comparisons between IMPT-CONV, IMPT-RF and FLASH-5Gy. 2D dose distributions for (d) IMPT-CONV and (e) FLASH-5Gy. (f) DVH comparisons between IMPT-CONV, FLASH-3Gy and FLASH-1Gy. The PTV-GTV, liver-GTV, duodenum, stomach and skin are contoured in blue, orange, green, purple and pink, respectively. The dose scale in isodose plot ranges from 10% to 110% of the prescription dose.

## 4. Discussion

In this study, we validated the concept of modularizing streamlined pin-RFs within a single-energy proton FLASH planning framework for liver SABR. While the pin-RFs developed using our framework are characterized by a coarser resolution and fewer steps along each ridge pin than existing pin-RF designs(Liu *et al* 2023, Zhang *et al* 2022a)—an attribute that already simplifies the production process, including 3D printing—the modularization technique we have introduced further refines production efficiency. By modularizing these pin-RFs, we have established a more streamlined manufacturing approach through the assembly of fixed-size modules from a pre-defined set. These reusable modules could not only reduce the production cost, but also facilitate easy modifications of the pin-RFs during the treatment. Moreover, we adopted graphite for range shifters due to its higher RSP and lower MCS effect compared to PMMA, potentially enabling smaller proton spot sizes and thus enhancing the dose conformity of the single-energy FLASH plan.

Our analysis of modularized pin-RFs demonstrated a primary demand for modules of 6 mm widths, followed by 2 to 4 mm widths, with minimal need for 1- and 5-mm modules. This distribution informs an optimal module set composition. Furthermore, our findings revealed a correlation between the complexity of energy modulation in IMPT-RF planning and factors such as target size and the number of beams. Larger targets, as illustrated in Case 3, required a greater number of ridge pins and exhibited more complex energy modulation. In contrast, the smaller target in Case 1 also demanded intricate energy modulation, suggesting that fewer beams necessitate increased complexity per beam to ensure optimal dose delivery. This higher energy modulation complexity was further echoed in the demand for thinner steps in Cases 1 and 3, compared to Case 2, which suggested finer energy adjustments for optimal dose delivery. Thus, extending the application of this method to additional cases and varied treatment sites could further refine our understanding of modularized pin-RFs and lead to a more optimal configuration of unit modules for the predefined set. Besides, our FLASH planning framework enables the adjustment of parameters, including energy layer spacing and the criteria for spot removal in the spot reduction process during IMPT-DS planning. This flexibility allows for the fine-tuning of pin and step counts, thus facilitating the customization of the resolution of the pin-RFs. Consequently, it opens the possibility for exploring unit modules of various sizes and resolutions, enhancing the assembly process of pin-RFs.

It should be noted that water was employed in this study to simulate the unit modules for pin-RF assembly, selected for its compatibility with the highest available resolution for dose calculation in RayStation 10B, which is 1 mm voxel size. This resolution allows module thicknesses to precisely match the required WETs in integer mm, ensuring alignment with the dose grid when assembling pin-RFs. This alignment is crucial for maintaining desired pin-RF shapes and achieving accurate dose calculations. Notably, materials with a low atomic number and minimal MCS effects are preferred for range shifters to enhance FLASH plan quality, a principle applicable to pin-RFs as well. To optimize the quality of FLASH plans, it is beneficial to explore alternative materials, along with adjusting sizes and resolutions of the unit modules for pin-RF assembly. Such investigations can be conducted by advanced Monte Carlo dose calculation tools (e.g., Geant4) allowing high resolution simulations and can be further supported by experimental measurements.

The FLASH doses in single-energy IMPT-RF plans were assessed using a phenomenological FLASH effectiveness model. This model integrates considerations of the dose threshold, FLASH persistence, and the dose rate threshold necessary for FLASH activation, providing a comprehensive reflection of the FLASH effect. The quantification of FLASH effects across all cases, in terms of physical dose reductions and clinical benefits at various dose thresholds for FLASH triggering, confirmed the strong correlations between dose threshold and the FLASH effect as previously demonstrated. These findings further provide insights into how the FLASH effect correlates with factors such as beam dose, beam arrangement, and the location of normal tissues. Notably, lower fractional beam doses necessitate a reduced dose threshold for FLASH triggering, typically resulting in a diminished FLASH effect. This pattern is evident in Cases 2 and 3, which involve three beams, compared to Case 1 with two beams. Additionally, closer beam arrangements, observed in Case 3, lead to decreased fractional beam doses, subsequently reducing the FLASH effect. The employment of more beams, particularly with additional pin-RFs incorporated in FLASH planning, increases low-dose exposure, causing more extensive low-dose spillage. Consequently, normal tissues located in areas exposed to this spillage often exhibit reduced benefits from the FLASH effect and may even experience negative clinical benefits. It is crucial to recognize that lowering the dose threshold to zero does not enable full dose delivery as FLASH to all normal tissue voxels, especially those receiving doses from multiple spot deliveries, due to the insufficient dose rate for FLASH trigger under these conditions.

Based on the correlations revealed between the FLASH effect and factors such as beam dose, beam arrangement, and the location of normal tissues, several considerations are essential in planning to maximize the FLASH effect. First, protocols that prescribe large fractional doses are essential, as they ensure the high fractional beam doses needed for a significant FLASH effect. Second, it is vital to minimize the number of beams while still ensuring adequate target dose coverage and sparing of normal tissues, in order to maintain substantial fractional beam doses. Third, reducing beam intersections beyond the target area, where feasible, can help increase spot doses critical for triggering the FLASH effect. Besides, normal tissues affected by low-dose spillage would benefit from a narrower beam penumbra, achievable through the use of collimators in conjunction with range shifters. Further applications of our FLASH planning method across various cases and treatment sites are necessary to refine and validate these strategies.

## 5. Conclusions

In this study, we demonstrated the feasibility of modularizing the streamlined pin-RFs in single-energy proton FLASH planning for liver SABR. Our analysis of modularized pin-RFs provides guidance on the optimal composition of the module set. Additionally, our findings from quantifying FLASH effects across clinical cases offer valuable insights into the relationships between the FLASH effect and factors such as beam dose, beam arrangement, and the location of normal tissues. These insights are critical for enhancing planning strategies to maximize the FLASH effect. It is crucial to apply the proposed method to additional cases and varied treatment sites to optimize the configuration of the unit module set for enhance pin-RF assembly and to refine planning strategies to maximize the FLASH effect.

**Conflicts of interest**

The authors have no conflicts to disclose.

**Acknowledgment**

This research is supported in part by the National Institutes of Health under Award Number R01CA272991, R56EB033332, R01EB032680, and U54CA274513.